\documentclass[twocolumn,amsmath,amssymb,showpacs,pre,superscriptaddress]
{revtex4}
\usepackage[utf8]{inputenc}
\usepackage{color,graphicx,subfigure,t1enc,grffile}

\newcommand{\eqnref}[1]{Eq.~\ref{eq:#1}}
\newcommand{\picref}[1]{Fig.~\ref{fig:#1}}
\newcommand{\chref}[1]{\ref{ch:#1}}

\newcommand{\RA}[1]{#1}
\newcommand{\xRA}[1]{}
\newcommand{\RB}[1]{#1}
\newcommand{\RAB}[1]{#1}
\newcommand{\RX}[1]{#1}

\bibstyle{apsrev}

\begin{document}

\title[short]{A simplified particulate model for coarse-grained
  hemodynamics simulations}

\author{F. Janoschek}
\email{fjanoschek@tue.nl}
\affiliation{Department of Applied Physics, Eindhoven University of
  Technology, P.\,O. Box 513, 5600\,MB Eindhoven, The Netherlands}
\affiliation{Institute for Computational Physics, University of Stuttgart,
  Pfaffenwaldring 27, 70569 Stuttgart, Germany}
\author{F. Toschi}
\email{f.toschi@tue.nl}
\affiliation{Department of Applied Physics, Eindhoven University of
  Technology, P.\,O. Box 513, 5600\,MB Eindhoven, The Netherlands}
\affiliation{Department of Mathematics and Computer Science, Eindhoven
  University of Technology, P.\,O. Box 513, 5600\,MB Eindhoven, The Netherlands}
\author{J. Harting}
\email{j.harting@tue.nl}
\affiliation{Department of Applied Physics, Eindhoven University of
  Technology, P.\,O. Box 513, 5600\,MB Eindhoven, The Netherlands}
\affiliation{Institute for Computational Physics, University of Stuttgart,
  Pfaffenwaldring 27, 70569 Stuttgart, Germany}

\date{\today}

\begin{abstract}
  \RA{Human blood flow is a multi-scale problem: in first
    approximation, blood is a dense suspension of plasma and
    deformable red cells. Physiological vessel diameters range from
    about one to thousands of cell radii.
    Current computational models either involve a homogeneous fluid
    and cannot track particulate effects or describe a relatively
    small number of cells with high resolution, but are incapable to
    reach relevant time and length scales.
    Our approach is to simplify much further than existing particulate
    models. We combine well established methods from other areas of
    physics in order to find the essential ingredients for a
    minimalist description that still recovers hemorheology. These
    ingredients are a lattice Boltzmann method describing rigid
    particle suspensions to account for hydrodynamic long range
    interactions and---in order to describe the more complex
    short-range behavior of cells---anisotropic model potentials known
    from molecular dynamics simulations. Paying detailedness, we
    achieve an efficient and scalable implementation which is crucial
    for our ultimate goal: establishing a link between the collective
    behavior of millions of cells and the macroscopic properties of
    blood in realistic flow situations. In this paper we present our
    model and demonstrate its applicability to conditions typical for
    the microvasculature.}
\end{abstract}
\maketitle

\section{Introduction}

Human blood is not a homogeneous substance but can be approximated as
a suspension of deformable red blood cells (RBCs, also called
erythrocytes) in a Newtonian liquid, the blood plasma. Under
physiological conditions the volume concentration of RBCs typically
amounts to $40\,\%$ to $50\,\%$ in larger blood vessels. We neglect
the other constituents like leukocytes and thrombocytes in this work
due to their far lower volume concentrations~\cite{goldsmith75}. In
the absence of external stresses, erythrocytes assume the shape of
biconcave discs with a diameter of approximately
$8\,\mu\text{m}$~\cite{evans72}. Their main biological task is the
transport of oxygen in the body, but due to the high volume fraction
they also strongly \RX{a}ffect the rheology of blood and its clotting
behavior~\cite{fung81}. An understanding of these effects is necessary
in order to study and to cure pathologically deviating phenomena in
the body and to design microfluidic devices for improved blood
analysis. In both cases, blood often has to be studied within complex
geometries that elude an analytical description. However, also the
computational treatment of blood is demanding. On large scales like in
arteries with diameters in the order of millimeters, blood can be
modeled as a continuous and even Newtonian fluid~\cite{boyd07}. Even
then, the computational effort and the complexity of the model can be
significant if realistic geometries show features which stretch over
different length scales.

For modeling flow in the microvascular network, there is need for a
description that accounts for the presence of discrete
cells~\cite{popel05}. Recently, models of deformable cells were
presented amongst others by Dupin et al.\ \cite{dupin08}, in the group
of Gompper~\cite{mcwhirter09}, and by Wu and Aidun~\cite{wu10}. Here,
the cell membrane is \RA{simplified to} a deformable mesh and coupled
to a mesoscopic simulation method for the plasma like multi-particle
collision dynamics~\cite{mcwhirter09} or lattice Boltzmann
(LB)~\cite{dupin08,wu10}. However, mainly because of the high
resolution that is necessary for the elaborate description of the
cells, these models are computationally too demanding for the
application to considerably larger 3D systems.

Our motivation is to bridge the scales that are accessible with both
classes of existing models by an intermediate approach: we keep the
particulate nature of blood, but try to \RA{find a minimal description
  of each single cell. We thus deliberately simplify much further than
  the authors of the particulate models cited above in order to gain
  the potential for a computationally efficient and scalable
  implementation. Resorting to well-established methods from other
  areas of physics we explore the ingredients necessary to recover the
  rheological behavior of blood. It is not our motivation to account
  for sub-cell effects in more than a coarse-grained way. In this work
  we aim at the description and validation of such a model while
  presenting possible applications in the range from approximately
  $10\,\mu\text{m}$ upward. Our} ultimate goal is to develop a
quantitative method that allows to study the flow in realistic
geometries but also to link bulk properties, for example the apparent
viscosity, to phenomena at the level of single erythrocytes. In the
case of cell deformation and aggregation in plane shear flow, this
link has been established already in experiment and
theory~\cite{chien70}. Numerical simulations enable us to extend this
knowledge to the case of arbitrary geometries and time-dependent
flows. Further microscopic properties of interest are the alignment of
cells or local changes of the cell concentration. The improved
understanding of the dynamic behavior of blood might be used for the
optimization of macroscopic simulation methods. Only a computationally
efficient description allows the accumulation of firm statistical data
that is necessary for this task.

The main idea of our model is to distinguish between the long-range
hydrodynamic coupling of cells and the short-range interactions that
are related to the complex mechanics, electrostatics, and the
chemistry of the membranes. Long-range hydrodynamic interactions are
considered by means of the LB method~\cite{succi01}. This mesoscopic
simulation method allows a relatively easy implementation of complex
boundary conditions which are needed for the simulation of realistic
geometries. Moreover, an efficient parallelization is straightforward
which even with our simplified model is crucial for the accumulation
of statistically relevant data or for the description of realistic
systems like branching vessels. Research on a parallel and efficient
implementation of the LB method for the simulation of flow in sparse
vessel networks was published by various
authors~\cite{axner08,mazzeo08,bernaschi09}. Consequently, the LB
method was applied to blood flow already in earlier works though they
differ from our approach in the accuracy with which cells are
resolved. For example, Boyd et al.\ \cite{boyd07} model blood either
as a Newtonian fluid or a homogeneous fluid with a shear
rate-dependent viscosity. Further, the studies by Dupin et al.\
\cite{dupin08}, Wu and Aidun~\cite{wu10}, Migliorini et al.\
\cite{migliorini02}, and by Sun and Munn~\cite{sun06} involve LB
solvers, but describe each RBC as either deformable or equipped with
an elaborate cell adhesion model.

In contrast to those, we are interested in a minimal resolution of
RBCs since reducing the resolution generally is the most effective way
to enhance the efficiency of a fluid dynamics solver. Ahlrichs and
D\"unweg implement a dissipative coupling of point-particles to an LB
fluid~\cite{ahlrichs98}. However, the description of RBCs as point
particles would involve a resolution which is so low that
hydrodynamics in the smallest vessels becomes
inaccessible. \RB{Concerning hydrodynamics, it is questionable whether
  the resolution of cell deformation has a benefit compared to a rigid
  particle model if each RBC is resolved with only a few lattice
  spacings.} In consequence we decide for a method for suspensions of
rigid particles of finite size that is similar to the one described
in~\cite{aidun98}. \RA{As will be explained below, not
  volume-conserving cell deformations of the order of one lattice
  spacing occur as an artefact of the method already but do not show
  significant influence on the flow behavior.} \RA{Ding and Aidun
  simulated rigid particles with the biconcave shape of unstressed red
  blood cells using an LB method~\cite{ding06}. It is known, however,
  that RBCs abandon this equilibrium shape and instead resemble
  elongated ellipsoids when exposed to shear
  flow~\cite{fischer78}. Thus, taking} into account the limited
lattice resolution, \RA{we decide for discretized ellipsoidal model
  cells with rotational symmetry as an approximation of} the shapes
actually assumed by real erythrocytes in many flow
situations. Differently from \RA{\cite{nguyen02,ding03,ding06}}, our
implementation does not enforce rigid particle surfaces since this
would be in contradiction with the nature of deformable
erythrocytes. \RA{Especially in bulk flow at high volume
  concentrations but also in capillaries due to the influence of
  walls, the flow is not dominated by long-range hydrodynamics but by
  short-range cell-cell and cell-wall effects. Thus, a coarse-grained
  description using effective cell and wall interactions is
  appropriate. We} account for the complex short-range behavior of
RBCs on a phenomenological level by means of model
potentials. \RAB{Our potentials serve to provide a softly repulsive
  core that follows the approximated ellipsoidal RBC shape. For this
  purpose}, the \RAB{method} of Berne and Pechukas~\cite{berne72} is
\RAB{applied} in order to \RAB{anisotropically} rescale a Hookian
spring \RAB{potential.}

In the following section \chref{hydrodynamics} we provide an
introduction to the application of the LB method to suspensions of rigid
particles and discuss how our model differs from other implementations. In
section \chref{potentials} we develop phenomenological potentials for the
anisotropic interaction of two cells and of cells and walls. Section
\chref{results} opens with the search for a parametrization which fits to
experimental literature data. This set of parameters is then used to
demonstrate the applicability of the new model to confined systems. We
further discuss the performance of our implementation for large systems
and conclude with a summary in section \chref{conclusion}.

\section{Hydrodynamic part of the model}\label{ch:hydrodynamics}

To model the blood plasma that surrounds the cells the LB method is
applied. For a comprehensive introduction we refer to~\cite{succi01}. The
fluid traveling with one of $r$ discrete velocities $\mathbf{c}_r$ at the
three-dimensional lattice position $\mathbf{x}$ and discrete time $t$ is
resembled by the single particle distribution function
$n_r(\mathbf{x},t)$. Its evolution in time is determined by collision
\begin{equation}
  \label{eq:col}
  n^*_r(\mathbf{x},t)
  =
  n_r(\mathbf{x},t)
  -
  \Omega
\end{equation}
and the successive advection
\begin{equation}
  \label{eq:adv}
  n_r(\mathbf{x}+\mathbf{c}_r,t+1)
  =
  n^*_r(\mathbf{x},t)
\end{equation}
of the post-collision distribution $n^*_r(\mathbf{x},t)$. \eqnref{col}
and \eqnref{adv} together can be written as the lattice Boltzmann
equation
\begin{equation}\label{eq:lbe}
  n_r(\mathbf{x}+\mathbf{c}_r,t+1)
  =
  n_r(\mathbf{x},t)
  -
  \Omega
  \text{ .}
\end{equation}
For the sake of simplicity and computational efficiency we follow a
D3Q19 approach with a single relaxation time $\tau$~\cite{qian92}. We
thus have $19$ discrete velocities and the BGK-collision
term~\cite{bhatnagar54}
\begin{equation}
  \Omega
  =
  \frac
  {n_r(\mathbf{x},t)-
    n_r^\text{eq}(\rho(\mathbf{x},t),\mathbf{u}(\mathbf{x},t))}
  {\tau},
\end{equation}
where the equilibrium distribution function
\begin{eqnarray}
  n_r^\text{eq}(\rho,\mathbf{u})
  & = &
  \rho\alpha_{c_r}
  \Bigg[
  1
  +\frac{\mathbf{c}_r\mathbf{u}}{c_\text{s}^2}
  +\frac{\left(\mathbf{c}_r\mathbf{u}\right)^2}{2c_\text{s}^4}
  -\frac{\mathbf{u}^2}{2c_\text{s}^2}\nonumber\\
  & & {}+\frac{\left(\mathbf{c}_r\mathbf{u}\right)^3}{6c_\text{s}^6}
  -\frac{\mathbf{u}^2\mathbf{c}_r\mathbf{u}}{2c_\text{s}^4}
  \Bigg]\label{eq:neq}
\end{eqnarray}
is an expansion of the Maxwell-Boltzmann distribution of third order
in velocity $\mathbf{u}$~\cite{chen00}. The value of the speed of
sound $c_\text{s}=1/\sqrt{3}$ depends on the choice of the
lattice. The same holds for the lattice weights
\begin{equation}
  \alpha_{c_r}
  =
  \left\{
    \begin{array}{l@{\quad\text{for }c_r=\,\,}l}
      1/3 & 0\\
      1/18 & 1\\
      1/36 & \sqrt{2}
    \end{array}
  \right.
  \text{ ,}
\end{equation}
which differ for lattice velocities $\mathbf{c}_r$ according to their
absolute value $c_r$. The local density
\begin{equation}
  \rho(\mathbf{x},t)=\sum_rn_r(\mathbf{x},t)
\end{equation}
and velocity
\begin{equation}
  \mathbf{u}(\mathbf{x},t)=
  \frac{\sum_rn_r(\mathbf{x},t)\mathbf{c}_r}{\rho(\mathbf{x},t)}
\end{equation}
are calculated as moments of the fluid distribution with respect to
the set of discrete velocities. Both are invariants of the BGK
collision rule \eqnref{col}. This method is well established for the
simulation of the liquid phase of suspensions~\cite{aidun98,nguyen02},
namely blood~\cite{dupin08,wu10}. It can be shown that in the limit of
small velocities and lattice spacings the Navier-Stokes equations are
recovered with a kinematic viscosity $\nu=(2\tau-1)/6$.

For a coarse-grained description of the hydrodynamic interaction of
cells and blood plasma, a method similar to the ones explained in
\cite{aidun98} and \cite{nguyen02} modeling rigid particles of finite
size is applied. Starting point is the mid-link bounce-back boundary
condition that implements no-slip boundaries for the fluid:
arbitrarily shaped geometries are discretized on the lattice by
turning the lattice nodes on the solid side of the theoretical
solid-fluid interface into fluid-less wall nodes. If $\mathbf{x}$ is
such a node the updated distribution at $\mathbf{x}+\mathbf{c}_r$ is
not determined by the advection rule \eqnref{adv} but according to
\begin{equation}\label{eq:lbe-bb}
  n_r(\mathbf{x}+\mathbf{c}_r,t+1)
  =
  n^*_{\bar{r}}(\mathbf{x}+\mathbf{c}_r,t)
\end{equation}
which means replacing the local distribution with the post-collision
distribution of the opposite direction $\bar{r}$ (defined by
$\mathbf{c}_{\bar{r}}\equiv-\mathbf{c}_r$). We make use of this
boundary condition to implement (rigid) vessel walls.

To model boundaries moving with velocity $\mathbf{v}$, \eqnref{lbe-bb}
can be complemented with a correction term
\begin{equation}\label{eq:c}
  C
  =
  \frac
  {2\alpha_{c_r}}
  {c_\text{s}^2}
  \rho(\mathbf{x}+\mathbf{c}_r,t)
  \,
  \mathbf{c}_r\mathbf{v}
\end{equation}
which is of first order in velocity. Inserting \eqnref{neq} and
\eqnref{c}, one can easily prove that the new update rule
\begin{equation}\label{eq:lbe-bb-corr}
  n_r(\mathbf{x}+\mathbf{c}_r,t+1)
  =
  n^*_{\bar{r}}(\mathbf{x}+\mathbf{c}_r,t)
  +
  C
\end{equation}
is up to second order consistent with the equilibrium distribution
function \eqnref{neq} for the general case
$\mathbf{u}=\mathbf{v}\not=\mathbf{0}$.

When used to implement freely moving particles, it is necessary to
keep track of the momentum
\begin{equation}\label{eq:deltap}
  \Delta\mathbf{p}_\text{fp}
  =
  \left(
    2n_{\bar{r}}
    +
    C
  \right)
  \mathbf{c}_{\bar{r}}
  \text{ ,}
\end{equation}
which is transferred during each time step by each single bounce-back
process. According to the choice of unit time steps it is equal to the
resulting force on the particle. The equations of motion of the
particles are integrated like in classical molecular dynamics (MD)
simulations to achieve the time evolution of the system. We implement
a combined LB/MD code in which both the time step and the spatial
decomposition scheme are shared between the two methods. A leap-frog
integrator that is adapted to the internal representation of the cell
orientations based on unit quaternions is applied \cite{allen91}.

Due to discretization errors the representation of a particle on the
lattice slightly changes its shape and volume during movement. When
new lattice sites are covered, the fluid at those is deleted. When a site
formerly occupied by a particle is freed, new fluid is created
according to \eqnref{neq}. In doing so, the initial density
$\bar{\rho}$ of the simulation is utilized as $\rho$. The velocity
$\mathbf{u}$ is estimated according to the translational and
rotational velocity of the particle and the no-slip assumption. In
both cases the change in total fluid momentum is balanced by an
additional force on the respective particle.

Physiological RBCs, however, are deformable and assume the shape of
biconcave discs in the absence of external stresses~\cite{evans72}.
Despite the coarse-graining in our model, we do not want to give up
the anisotropy of RBCs. Obviously, anisotropic model cells are able to
display a much richer behavior than radially symmetric particles. We
thus choose a simplified ellipsoidal geometry that is defined by two
distinct half-axes $R_\parallel$ and $R_\perp$ parallel and
perpendicular to the unit vector $\hat{\mathbf{o}}_i$ which points
along the direction of the axis of rotational symmetry of each
particle $i$.

Closely approaching particles are modeled as follows: when
there are still fluid nodes between both discretized volumes the LB
method is able to keep track of the emerging lubrication forces apart
from discretization errors. As soon as there is a direct
particle-particle interface without intermediate fluid nodes, the
lubrication forces cannot be covered by a lattice-based method
anymore. Moreover, an effective attraction becomes visible because of
the missing fluid pressure in between the particles. Typical
applications of this simulation method to the case of dense
suspensions additionally feature analytical short-range lubrication
corrections to overcome this problem~\cite{ding03,nguyen02}. These are
implemented as pair-forces that depend on the relative velocity and
diverge for vanishing gap-widths. However, this procedure is
inappropriate for a model for suspensions of deformable cells. Since
the theoretical particle shapes defined by $R_\parallel$ and $R_\perp$
are fixed, our application even requires tolerance for the overlap of
the discretized volumes in order to account for the case where two
cells strongly deform while approaching each other. Due to the
complexity of the emerging forces that include electrostatic repulsion
and van der Waals forces but also the mechanics and chemistry of the
cell membranes and the rheology of the cell plasma, we cover them on a
purely phenomenological level in the next section. Here, we support
the LB method with additional rules which result in forces for the
case of two particles in direct contact with each other that are
neither divergent nor excessively attractive. Wherever a direct
particle-particle interface is encountered, we apply a pair of mutual
forces
\begin{equation}\label{eq:deltapppp}
  \mathbf{F}_\text{pp}^+
  =
  2n_r^\text{eq}(\bar{\rho},\mathbf{u}=\mathbf{0})\mathbf{c}_r
\end{equation}
and
\begin{equation}\label{eq:deltapppm}
  \mathbf{F}_\text{pp}^-
  =
  2n_{\bar{r}}^\text{eq}(\bar{\rho},\mathbf{u}=\mathbf{0})\mathbf{c}_{\bar{r}}
  =
  -\mathbf{F}_\text{pp}^+
\end{equation}
at each link across the interface which are directed towards each
respective particle. Comparison with \eqnref{deltap} shows that this
is exactly the momentum transfer during one time step due to the
rigid-particle algorithm for a resting particle and an adjacent site
with resting fluid at equilibrium and initial density
$\bar{\rho}$. The fluid in our simulation is to good approximation
incompressible and the velocities are small. The forces arising from
those regions of the particle surfaces that are in contact with the
fluid therefore are largely compensated and do not cause an artificial
attraction. In consequence, the self-induced collapse of particles in
contact is prevented without the need for divergent lubrication
corrections as in rigid-particle models. Moreover, for a given system,
\eqnref{deltapppp} and \eqnref{deltapppm} depend only on
$\mathbf{c}_r=-\mathbf{c}_{\bar{r}}$. For symmetry reasons,
$n_r^\text{eq}(\bar{\rho},\mathbf{0})=n_{\bar{r}}^\text{eq}(\bar{\rho},\mathbf{0})$
holds. Thus, the momentum balance is kept since the two forces
emerging from any particle-particle link compensate each
other. However, since \eqnref{deltapppp} and \eqnref{deltapppm} do not
depend on the relative velocity they cannot cover dissipative forces
between particles. This limitation needs to be kept in mind when
deciding about phenomenological cell-cell forces and their
parametrization later in this paper.

For the sake of simplicity we do not allow a lattice node to be
occupied by more than one cell. Occupation instead is determined by
the order in which cells arrive at a node. From the point of view of
the surrounding fluid this behavior is physically consistent with two
particles that compressibly deform upon contact. The compressibility,
however, can lead to an artificial increase of the total mass since
the number of fluid nodes increases temporarily and we do not adjust
the particle mass dynamically according to the volume occupied
momentarily. Thus, in an ensemble of many cells, there are
fluctuations of the total mass due to the introduction of the
correction term $C$ in \eqnref{lbe-bb-corr} and due to the change in
total volume of the solid phase which fluctuates when cells move and
increases when they overlap. However, even during millions of time
steps we find no drift of the total mass for the systems we simulate
here.

In case of close contact of cells with the confining geometry we
proceed in a similar manner as for two cells. The only difference is
that the forces on the system walls are ignored since they are assumed
to be rigid and fixed.

\section{Model potentials for cell-cell and cell-wall
  interactions}\label{ch:potentials}

In order to account for the complex behavior of real RBCs at small
distances we add phenomenological pair potentials. For simplicity, we
restrict ourselves to repulsive forces. This can be justified because
in many physiological situations of interest, for example close to the
walls of large parts of the arterial system, high shear rates render
aggregative effects negligible~\cite{stroev07,chien70}. \RAB{The task
  of the potential therefore lies in establishing an excluded volume
  for each cell. Due to the mild increase of the potential, an overlap
  of these volumes will be unfavorable yet possible to some
  degree. Thus, the deformation of cells upon contact is modeled in a
  phenomenological way.} A simplified potential also is
\RB{beneficial} to the efficiency of the model since it can be
evaluated with less numerical effort and is less likely to demand
small time steps or high order integrators. We start with the
repulsive branch of a Hookian spring potential
\begin{equation}\label{eq:phi}
  \phi(r_{ij})
  =
  \left\{
    \begin{array}{l@{\qquad}l}
      \varepsilon\left(1-r_{ij}/\sigma\right)^2 &
      r_{ij}<\sigma\\
      0 &
      r_{ij}\ge\sigma
    \end{array}
  \right.
\end{equation}
for the scalar displacement $r_{ij}$ of two particles $i$ and
$j$. This is probably the simplest way to describe (elastic)
deformability. The energy at zero displacement and the distance at
which the repulsive potential force sets in can be directly controlled
by means of the parameters $\varepsilon$ and $\sigma$. With respect to
the disc-like shape of RBCs, we follow the approach of Berne and
Pechukas~\cite{berne72} and choose the stiffness parameter
\begin{equation}\label{eq:epsilon}
  \varepsilon(\hat{\mathbf{o}}_i,\hat{\mathbf{o}}_j)
  =
  \frac
  {\bar{\varepsilon}}
  {\sqrt{1-\chi^2\left(\hat{\mathbf{o}}_i\hat{\mathbf{o}}_j\right)^2}}
\end{equation}
and the size parameter
\begin{equation}\label{eq:sigma}
  \sigma(\hat{\mathbf{o}}_i,\hat{\mathbf{o}}_j,\hat{\mathbf{r}}_{ij})
  =
  \frac
  {\bar{\sigma}}
  {\sqrt{1-\frac{\chi}{2}\left[
        \frac
        {\left(
            \hat{\mathbf{r}}_{ij}\hat{\mathbf{o}}_i
            +
            \hat{\mathbf{r}}_{ij}\hat{\mathbf{o}}_j
          \right)^2}
        {1+\chi\hat{\mathbf{o}}_i\hat{\mathbf{o}}_j}
        +
        \frac
        {\left(
            \hat{\mathbf{r}}_{ij}\hat{\mathbf{o}}_i
            -
            \hat{\mathbf{r}}_{ij}\hat{\mathbf{o}}_j
          \right)^2}
        {1-\chi\hat{\mathbf{o}}_i\hat{\mathbf{o}}_j}
      \right]}}
\end{equation}
as functions of the orientations $\hat{\mathbf{o}}_i$ and
$\hat{\mathbf{o}}_j$ of the cells and their normalized center
displacement $\hat{\mathbf{r}}_{ij}$. We achieve an anisotropic
potential with a zero-energy surface that is approximately that of
ellipsoidal discs. Their half-axes $\sigma_\parallel$ and
$\sigma_\perp$ parallel and perpendicular to the symmetry axis enter
\eqnref{epsilon} and \eqnref{sigma} via
\begin{equation}
  \bar{\sigma}=2\sigma_\perp
  \quad\text{and}\quad
  \chi
  =
  \frac{\sigma_\parallel^2-\sigma_\perp^2}{\sigma_\parallel^2+\sigma_\perp^2}
  \text{ ,}
\end{equation}
whereas $\bar{\varepsilon}$ determines the potential strength. \RAB{The
  above approach for anisotropic rescaling of radial symmetric
  potentials and its later improvement by Gay and Berne \cite{gay81}
  were intended for modeling liquid crystal systems. Particularly the
  method by Gay and Berne is applied almost exclusively to a
  Lennard-Jones potential featuring a short range repulsion and an
  attraction on moderate distances. This is referred to as ``Gay-Berne
  potential'' in the literature. The model potential presented by us
  lacks the attractive tail but is equipped only with a softly
  repulsive core. In consequence, there is no force acting on
  particles separated by more than the respective core diameter and at
  physiological volume concentrations we cannot expect to observe
  spontaneous ordering of the system. Compared to typical liquid
  crystal applications, the role of our potential lies rather in
  providing a soft repulsion within an anisotropic discoid volume than
  in making specific cell alignments more favorable compared to
  others. \picref{potential-force} displays in dimensionless form the
  magnitude of the resultant repulsive pair force $F$ as a function of
  $r_{ij}$ for $\sigma_\perp=3\sigma_\parallel$ and three simple sets
  of relative orientations:
  $\hat{\mathbf{o}}_i\parallel\hat{\mathbf{o}}_j\perp\hat{\mathbf{r}}_{ij}$,
  $\hat{\mathbf{o}}_i\parallel\hat{\mathbf{o}}_j\parallel\hat{\mathbf{r}}_{ij}$,
  and
  $\hat{\mathbf{o}}_i\perp\hat{\mathbf{o}}_j\parallel\hat{\mathbf{r}}_{ij}$.
  Depending on the orientations, the repulsive force sets in at
  different $r_{ij}$. Aiming at the presentation of a model potential
  which is simplified to the greatest possible extent, we chose the
  Berne-Pechukas approach which is slightly less complex than the more
  popular one by Gay and Berne. In this approach, it is not possible
  to independently adjust the interaction strength for different
  molecular orientations. That the potential is considerably stiffer
  in the case where the flat sides of both cells are aligned towards
  each other is, however, consistent with the fact that for this
  orientation, the same linear approach creates a significantly larger
  overlap volume than in the other two cases. In the following
  section, we will find values for $\bar{\varepsilon}$,
  $\sigma_\parallel$, and $\sigma_\perp$ that reproduce the
  rheological behavior of blood.}

\begin{figure}
  \centering
  \RAB{\includegraphics[width=\columnwidth]
    {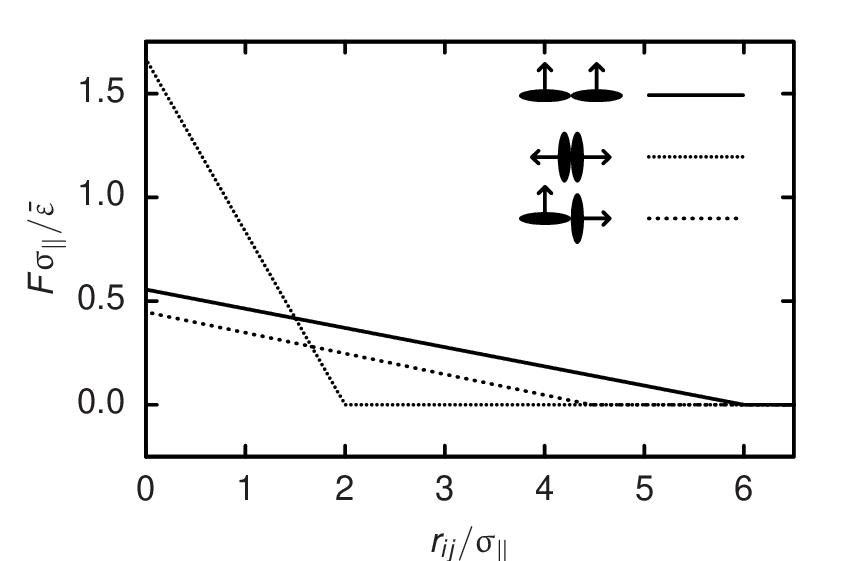}}
  \caption{\label{fig:potential-force}\RAB{Dimensionless repulsive
      potential force as a function of the dimensionless center
      distance for $\sigma_\perp=3\sigma_\parallel$ and three sets of
      relative orientations
      $\hat{\mathbf{o}}_i\parallel\hat{\mathbf{o}}_j\perp\hat{\mathbf{r}}_{ij}$,
      $\hat{\mathbf{o}}_i\parallel\hat{\mathbf{o}}_j\parallel\hat{\mathbf{r}}_{ij}$,
      and
      $\hat{\mathbf{o}}_i\perp\hat{\mathbf{o}}_j\parallel\hat{\mathbf{r}}_{ij}$.
      An approximately ellipsoidal excluded volume can be deducted
      from the surface at which the repulsion sets in.}}
\end{figure}

For modeling the cell-wall interaction we assume a sphere with radius
$\sigma_\text{w}=1/2$ at every lattice node on the surface of a
vessel wall and implement similar potential forces as for the
cell-cell interaction based on the repulsive spring potential
\eqnref{phi}. Berne and Pechukas show that using
\begin{equation}\label{eq:sigmaw}
  \sigma(\hat{\mathbf{o}}_i,\hat{\mathbf{r}}_{i\mathbf{x}})
  =
  \frac
  {\bar{\sigma}_\text{w}}
  {\sqrt
    {1
      -
      \chi_\text{w}
      \left(\hat{\mathbf{r}}_{i\mathbf{x}}\hat{\mathbf{o}}_i\right)^2}}
\end{equation}
instead of \eqnref{sigma} as a size parameter with
\begin{equation}
  \bar{\sigma}_\text{w}
  =
  \sqrt{\sigma_\perp^2+\sigma_\text{w}^2}
  \qquad
  \text{and}
  \qquad
  \chi_\text{w}
  =
  \frac
  {\sigma_\parallel^2-\sigma_\perp^2}
  {\sigma_\parallel^2+\sigma_\text{w}^2}
\end{equation}
allows to scale a potential with radial symmetry to fit for the
description of the interaction of a sphere and an ellipsoidal disc
\cite{berne72}. $\hat{\mathbf{r}}_{i\mathbf{x}}$ is the normalized
center displacement of particle $i$ and a wall node $\mathbf{x}$. It
is not necessary to scale the stiffness parameter anisotropically,
instead we set
$\varepsilon(\hat{\mathbf{o}}_i,\hat{\mathbf{o}}_j)=\bar{\varepsilon}_\text{w}$
fixed and use $\bar{\varepsilon}_\text{w}$ to tune the potential
strength. The values of $\sigma_\parallel$ and $\sigma_\perp$ are kept
the same as for the cell-cell interaction.

\picref{model-2d} shows a conclusive outline of the model. Two cells
$i$ and $j$ surrounded by blood plasma and a section of a vessel wall
are displayed. To enhance the explanatory power of the drawing we
choose to restrict ourselves to the presentation of a cut parallel to
the axes of rotational symmetry of the cells. Thus, the RBCs are
visualized as two-dimensional ellipses instead of three-dimensional
ellipsoids. Depicted are the cell shapes defined by the zero-energy
surface of the cell-cell potential \eqnref{phi} with \eqnref{sigma}
that can be approximated by ellipsoids with the size parameters
$\sigma_\parallel$ and $\sigma_\perp$ as half axes. Also shown are the
spheres with radius $\sigma_\text{w}$ defined accordingly by the
cell-wall interaction \eqnref{phi} and \eqnref{sigmaw} which are
assumed at all wall nodes that are linked to a fluid node by one of
the lattice directions $\mathbf{c}_r$. While these spheres are
centered on the respective wall nodes, the cells are free to assume
continuous positions and orientations $\mathbf{o}_i$ and
$\mathbf{o}_j$. In consequence, also the center displacement vectors
$\mathbf{r}_{ij}$ and $\mathbf{r}_{i\mathbf{x}}$ between the cells and
between cell $i$ and an arbitrary wall node $\mathbf{x}$ are
continuous. Still, for the cell-plasma interaction an ellipsoidal
volume with half axes $R_\parallel$ and $R_\perp$ is discretized on
the underlying lattice. For clarity, this is drawn only for cell $j$.

\begin{figure}
  \centering
  \includegraphics[width=\columnwidth]{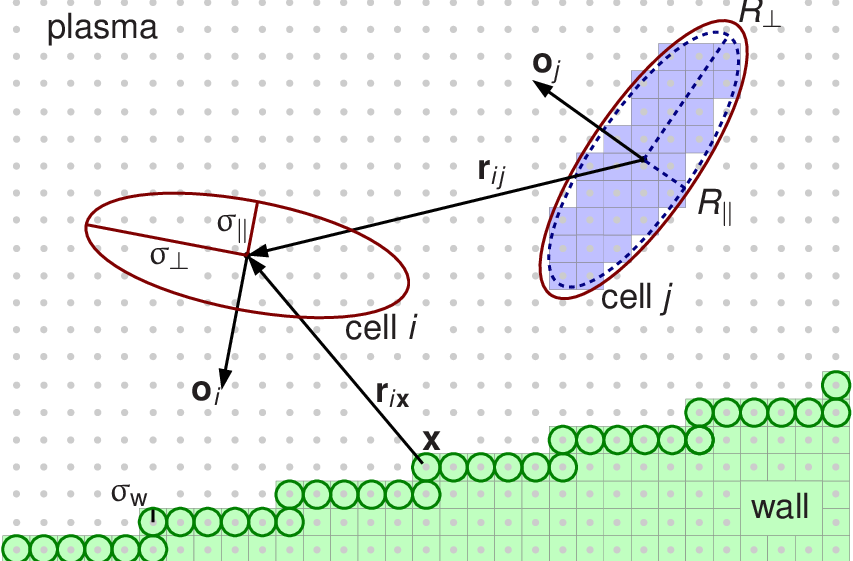}
  \caption{\label{fig:model-2d}(Color online) Outline of our 3D model
    by means of a
    2D cut. Shown are two cells $i$ and $j$ with their axes of
    rotational symmetry $\mathbf{o}_i$ and $\mathbf{o}_j$. The volumes
    defined by the cell-cell interaction is approximately ellipsoidal
    with half axes $\sigma_\parallel$ and $\sigma_\perp$ (red,
    ---). The ellipsoidal volume of the cell-plasma interaction with
    half axes $R_\parallel$ and $R_\perp$ is discretized on the
    underlying lattice. It is shown for only one cell (blue, \mbox{-
      -}). The cell-wall potential assumes spheres with radius
    $\sigma_\text{w}$ on all surface wall nodes (green, ---). Depicted
    are also the center displacement vectors $\mathbf{r}_{ij}$ and
    $\mathbf{r}_{i\mathbf{x}}$ between both cells and to an arbitrary
    surface wall node $\mathbf{x}$.}
\end{figure}

\section{Results}\label{ch:results}

As a convention in this work, primed variables are used to distinguish
quantities given in physical units from the corresponding unprimed variable
measured in lattice units. The maximum extent of physiological RBCs at
their equilibrium shape amounts to about $8\,\mu\text{m}$ and
$2.6\,\mu\text{m}$ perpendicular and parallel to the axis of
rotational symmetry~\cite{evans72}. We find that an ellipsoid of
revolution with the same numbers as axes has a volume of
$87\,\mu\text{m}^3$ which fits with the RBC volume measured
in~\cite{evans72}. We therefore choose the size parameters of the
cell-cell potential to be
\begin{equation}\label{eq:cc-dimension-parameters}
  \sigma'_\perp=4\,\mu\text{m}
  \quad\text{and}\quad
  \sigma'_\parallel=4/3\,\mu\text{m}
\end{equation}
and achieve that both the magnitude and the maximum extents of the
volume defined by the cell-cell interaction match typical values for
physiological erythrocytes.

All quantities that are of interest in our simulations can be
converted from simulation units to physical units by multiplication
with products of integer powers of the conversion factors $\delta x$,
$\delta t$, and $\delta m$ for space, time, and mass that thus
completely define a scale. We determine the mean deviations of the
Stokes drag coefficients of a single spherical particle from the
theoretical values in the laminar regime to be in the order of
$10^{-2}$ for a radius of $2.5$ lattice units. This is in agreement
with equivalent tests done by Ladd \cite{ladd94b}. We therefore
restrict ourselves to simulations of particles whose representation on
the lattice is at least as large as that of a sphere with radius
$2.5$. When using the same aspect ratio
$R_\perp/R_\parallel=\sigma_\perp/\sigma_\parallel=3$ for the
cell-fluid as for the cell-cell interaction, this requirement results
in minimum values for $R_\perp$ and $R_\parallel$ of $3.6$ and $1.2$
lattice units, respectively.

It can be expected that with cell-fluid volumes that are significantly
smaller than the size parameters of the potential we cannot achieve
realistic coupling strengths which are needed for example to model the
clogging of capillaries. Still, $R_\perp$ and $R_\parallel$ should be
smaller than the respective size parameters of the cell-cell potential
since limiting the amount of overlapping cell-fluid interaction volume
will improve the modeling of hydrodynamics between cells. Ladd et al.\
\cite{ladd01} suggest \RB{assisting} the particle-fluid coupling
method with lubrication corrections starting at gap widths below $2/3$
lattice spacings. \RB{Throughout this work, we} choose $\delta
x=2/3\,\mu\text{m}$ as a compromise that both keeps the resolution and
the computational cost low and allows \RB{one} to combine---for
example---a high ratio of $R_\parallel/\sigma_\parallel=7/8$ with a
minimum gap width of $2(\sigma_\parallel-R_\parallel)=0.5$ at which
the cell-cell potential starts to set in.

To improve the numerical stability of the LB method and to easily
relate given input radii $R_\parallel$ and $R_\perp$ to an effective
particle size\RA{~\cite{ladd94b}}, we always set the relaxation time
to $\tau=1$. This, together with the constraint
\begin{equation}\label{eq:dt-constraint}
  \nu
  \frac{\delta x^2}{\delta t}
  =
  \frac{2\tau-1}{6}
  \frac{\delta x^2}{\delta t}
  =
  1.09\times10^{-6}\frac{\text{m}^2}{\text{s}}
  =
  \nu'
\end{equation}
caused by the fact that the simulated kinematic fluid viscosity $\nu$
is supposed to match the kinematic viscosity of blood plasma of
$\nu'=1.09\times10^{-6}\,\text{m}^2/\text{s}$ when converted to
physical units, determines the time discretization as $\delta
t=6.80\times10^{-8}\,\text{s}$. For convenience, we arbitrarily
choose the fluid density in simulation units to be
$\bar{\rho}=1$. With $\delta x$ and the physical plasma density
$\bar{\rho}'=1.03\times10^3\,\text{kg}/\text{m}^3$, this choice
results in a mass conversion factor of $\delta
m=3.05\times10^{-16}\,\text{kg}$.

We first investigate the effects of the free model parameters by
measuring the ratio of the apparent dynamic viscosity $\mu_\text{app}$
and the constant plasma viscosity $\mu$ for a homogeneous suspension
of cells in plane Couette flow. All simulations reported here are
performed on a system with a size of $l_x=128$ lattice units in $x$-
and at least $l_y=l_z=40$ lattice units in $y$- and
$z$-direction. This represents $85\times27^2\,\mu\text{m}^3$ of real
blood. Between the two $yz$-side planes a constant offset of the local
fluid velocities in $z$-direction is imposed by an adaption of the
Lees-Edwards shear boundary condition to the LB
method~\cite{lees72,wagner99}. The other edges are linked purely
periodically. For the cells, we implement a reflective boundary
condition that negates the normal velocity component of a cell as soon
as its center distance to one of the sheared side planes becomes less
than $\sigma_\perp$. \RA{This procedure surely is inconsistent with
  respect to the open boundaries implemented for the fluid but far
  easier to achieve than a common Lees-Edwards implementation for both
  phases.} To \RA{prevent these boundaries from influencing} our
measurements, we determine the shear rate $\dot{\gamma}$ only in the
central half of the system \RA{where the flow resembles an unbounded
  Couette flow}. \RA{We obtain $\dot{\gamma}$ from a linear fit of}
the velocity profile $v_z(x)$. The apparent viscosity
\begin{equation}\label{eq:mu}
  \mu_\text{app}
  =
  \frac{\Delta p_{\text{LE},z}}{l_yl_z\dot{\gamma}}
\end{equation}
is then calculated based on $\Delta p_{\text{LE},z}$ which is the
averaged $z$-momentum transfer across both Lees-Edwards boundaries
during one time step. For each shear rate, we start with resting and
randomly oriented model cells suspended in likewise resting fluid. We
calculate \eqnref{mu} in intervals during the simulation and start
accumulating the result for temporal averaging as soon as a steady
state is achieved. Several samples prove that neither the change of
the random seed for the generation of the initial cell configuration
nor the stepwise increase of the system size perpendicular to the
velocity gradient up to a volume of $85^3\,\mu\text{m}^3$ leads to any
significant deviation of the results. However, we find that the shear causes
the cell orientations $\{\hat{\mathbf{o}}_i\}$ to preferably align in
the $xz$-plane.

A proper choice of the ratio $R_\parallel/\sigma_\parallel$ is not
known a priori. We thus perform simulations at a constant shear rate
of $\dot{\gamma}=(2.21\pm0.08)\times10^3\,\text{s}^{-1}$ for different
$R_\parallel/\sigma_\parallel$. \RB{The resulting particle Reynolds
  numbers $Re_\text{P}$ are of the order of $10^{-1}$.} We arbitrarily
choose a cell number density of
$p'=(6.4\pm0.3)\times10^{15}\,\mathrm{m}^{-3}$ corresponding to a
physiological hematocrit of $56\,\%$ and a cell stiffness parameter of
$\bar{\varepsilon}'=1.47\times10^{-15}\,\text{J}$. The resulting
apparent viscosity $\mu_\text{app}$ as a function of the ratio
$R_\parallel/\sigma_\parallel$ is drawn in
\picref{shear-ladd-volume}. A relatively mild and almost linear
increase is visible for $R_\parallel/\sigma_\parallel<1$ which can be
related to the increase of friction in the system. Around $1$, the
increase becomes considerably steeper as the minimum gaps of
approaching cells vanish. At even larger ratios, the slope decreases
again due to large and unphysical amounts of overlap of the cell-fluid
coupling volumes that accordingly to \eqnref{deltapppp} and
\eqnref{deltapppm} reduce the effective friction between cells. Based
on our previous considerations and affirmed by
\picref{shear-ladd-volume}, we choose
$R_\parallel/\sigma_\parallel=11/12\approx0.92$ as a value that is
close to unity but still induces an only moderate amount of overlap
even at high shear rates of the order of $10^3\,\text{s}^{-1}$. This
choice results in size parameters of the cell-fluid interaction of
\begin{equation}\label{eq:cf-dimension-parameters}
  R'_\perp=11/3\,\mu\text{m}
  \quad\text{and}\quad
  R'_\parallel=11/9\,\mu\text{m}
  \text{ .}
\end{equation}
All dimensions in \picref{model-2d} above were already drawn to scale
with respect to the dimensional parameters in
\eqnref{cc-dimension-parameters} and \eqnref{cf-dimension-parameters}.

\begin{figure}
  \centering
  \includegraphics[width=\columnwidth]
  {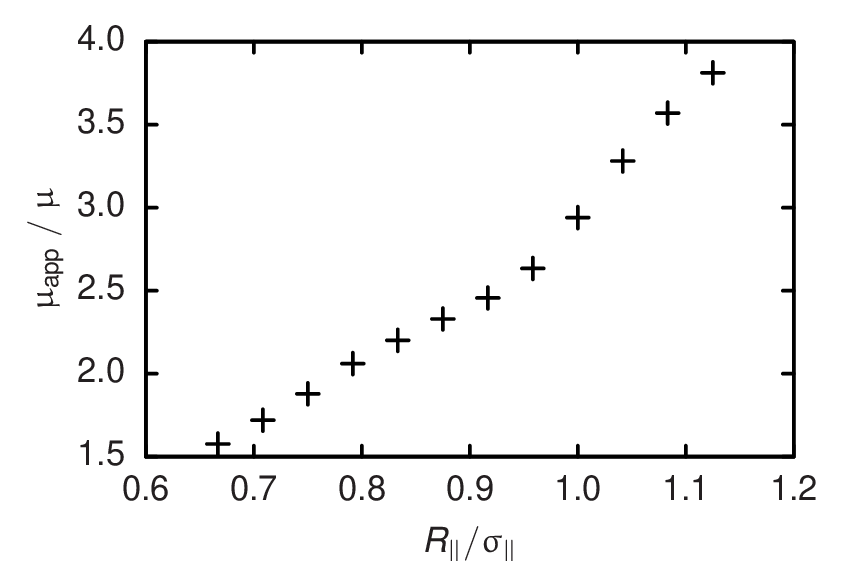}
  \caption{\label{fig:shear-ladd-volume}\RB{Dependence of the apparent
      dynamic viscosity $\mu_\text{app}$ on} the fraction
    $R_\parallel/\sigma_\parallel$ of the linear dimensions of the
    cell-fluid and cell-cell interaction volumes for a shear rate of
    $\dot{\gamma}'=(2.21\pm0.08)\times10^3\,\text{s}^{-1}$, a number
    density of $p'=(6.4\pm0.3)\times10^{15}\,\text{m}^{-3}$, a cell
    stiffness parameter
    $\bar{\varepsilon}'=1.47\times10^{-15}\,\text{J}$, and cell-cell
    size parameters $\sigma_\perp'=4\,\mu\text{m}$ and
    $\sigma_\parallel'=4/3\,\mu\text{m}$. All consecutive simulations
    are performed with
    $R_\parallel/\sigma_\parallel=11/12\approx0.92$.}
\end{figure}

The parameter $\bar{\varepsilon}$ is of special interest since it
controls the cell stiffness which describes the deformability of the
erythrocytes in our model. From experiments it is known that the shear
thinning behavior of blood at high shear rates is related to the
deformability of the RBC membrane and can be disabled by artificial
hardening of the cells~\cite{chien70,shin04}. Our implementation of
the model stays numerically stable only for a limited range of
$\bar{\varepsilon}$. Simulations performed for various shear rates
\RB{$1.7\times10^1\,\text{s}^{-1}<\dot{\gamma}<2.3\times10^4\,\text{s}^{-1}$
  corresponding to particle Reynolds numbers $10^{-3}\lesssim
  Re_\text{P}\lesssim1$} and $\bar{\varepsilon}'$ varying between
$1.47\times10^{-16}\,\mathrm{J}$ and $1.47\times10^{-12}\,\mathrm{J}$
at a cell-fluid volume concentration of $43\,\%$ still show that for a
given shear rate, larger $\bar{\varepsilon}$ result in higher
viscosities yet in a less steep viscosity
decrease. \picref{shear-epsilon} displays this effect which is
asymptotically consistent with the experimental results of
Chien~\cite{chien70}. It is interesting to note that by plotting the
apparent viscosity over the fraction
$\dot{\gamma}/\bar{\varepsilon}$---as we do in
\picref{shear-epsilon-scaled}---we can collapse the region of
strongest viscosity decrease in the curves for different
$\bar{\varepsilon}$. This indicates that the shear thinning is
determined by a balance of viscous and potential forces that scale
with $\dot{\gamma}$ and $\bar{\varepsilon}$, respectively. Comparison
of \picref{shear-epsilon} with experimental data taken from the
literature~\cite{chien70} shows best consistency in the case of high
shear rates $\dot{\gamma}'\sim10^3\,\mathrm{s}^{-1}$ for
$\bar{\varepsilon}'=1.47\times10^{-15}\,\mathrm{J}$. We keep this
value for all further simulations in the current work.

\begin{figure}
  \hspace*{\fill}
  \subfigure[]
  {\label{fig:shear-epsilon-unscaled}%
    \includegraphics[width=0.4\columnwidth]
    {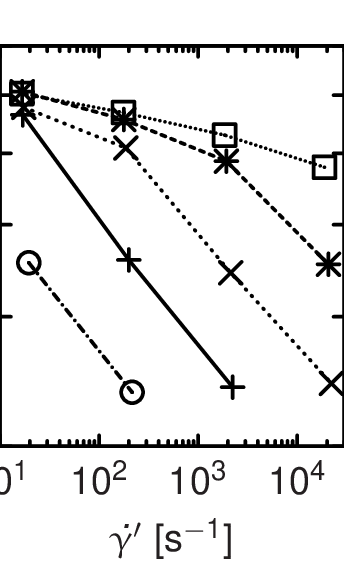}}
  \hspace*{0.25em}
  \subfigure[]
  {\label{fig:shear-epsilon-scaled}%
    \includegraphics[width=0.4\columnwidth]
    {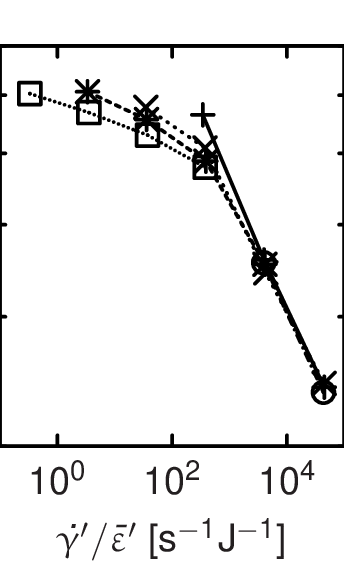}}
  \hspace*{0.75em}
  \caption{\label{fig:shear-epsilon}\subref{fig:shear-epsilon-unscaled}
    shows the shear rate dependent apparent viscosity $\mu_\text{app}$
    at a cell-fluid volume concentration of
    $\Phi_\text{cf}=43\,\%$. The different symbols represent different
    cell stiffness parameters $\bar{\varepsilon}'=1.47\times10^{-k}$
    with $k=16,15,14,13,12$ from bottom to top. Rescaling the shear
    rate $\dot{\gamma}'$ with $\bar{\varepsilon}'$ as displayed in
    \subref{fig:shear-epsilon-scaled} leads to a collapse of the
    region of steepest viscosity decrease on a single curve which
    hints at a concurrence of viscous and potential forces. All
    further simulations are performed with $k=15$.}
\end{figure}

Having defined values for all parameters of the cell-fluid and
cell-wall interaction, we can now investigate the effect of varying
cell concentrations on the viscosity. For given $R_\parallel$ and
$R_\perp$, the cell-fluid volume concentration $\Phi_\text{cf}$ is
proportional to the number
concentration. \picref{shear-volume-concentration} shows the
\RB{dependence of the apparent viscosity on} $\Phi_\text{cf}$ \RB{for
  a fixed shear rate of
  $\dot{\gamma}'=(2.2\pm0.1)\times10^3\,\text{s}^{-1}$. The particle
  Reynolds number is of the order of $Re_\text{P}\sim10^{-1}$.} For
$\Phi_\text{cf}<35\,\%$ we find a nearly linear increase of
$\mu_\text{app}$.  For $\Phi_\text{cf}>35\,\%$, the curve is still
linear but the slope is slightly smaller. Compared to the literature,
$\mu_\text{app}$ stays clearly below the viscosities known for hard
ellipsoids with a similar aspect ratio of $0.3$~\cite{bertevas10}. The
lower viscosities of our model, especially at high volume fractions,
are caused by the reduced dissipation between touching and overlapping
cells. This explanation can be substantiated by considering two
neighboring boxes of a periodic arrangement each containing one cell
in the center. Depending on the orientation and offset relative to the
LB grid, direct cell-cell links start to occur at volume
concentrations between $30\,\%$ and $50\,\%$ for the given $R_\perp$
and $R_\parallel$. These numbers also match the region in
\picref{shear-volume-concentration} where the slope of
$\mu_\text{app}(\Phi_\text{cf})$ decreases. As can be seen from
\picref{shear-volume-concentration}, our results for concentrations up
to about $50\,\%$ fit reasonably well with the experimental studies of
Goldsmith (see~\cite{fung81}) and of Shin et al.\ \cite{shin04}. At
higher $\Phi_\text{cf}$, touching cell-fluid volumes start to dominate
the rheology of the model suspension. The exact shear rates applied by
Goldsmith and Shin et al.\ are not known. We can only infer from the
literature that $\dot{\gamma}'$ was larger than $100\,\text{s}^{-1}$
and $250\,\text{s}^{-1}$, respectively. In this range, blood shows
shear thinning behavior~\cite{chien70,shin04} and so does our model
(see \picref{shear-epsilon}). It therefore is not possible to
determine whether---as \picref{shear-volume-concentration}
suggests---the model perfectly matches with experiments for
$\Phi_\text{cf}<40\,\%$ and underestimates the viscosity between
$40\,\%$ and $50\,\%$. However, \picref{shear-epsilon} demonstrates
that a better consistency at physiologically important concentrations
around $40\,\%$ should be easily attainable by tuning the value of
$\bar{\varepsilon}$.

\begin{figure}
  \centering
  \includegraphics[width=\columnwidth]
  {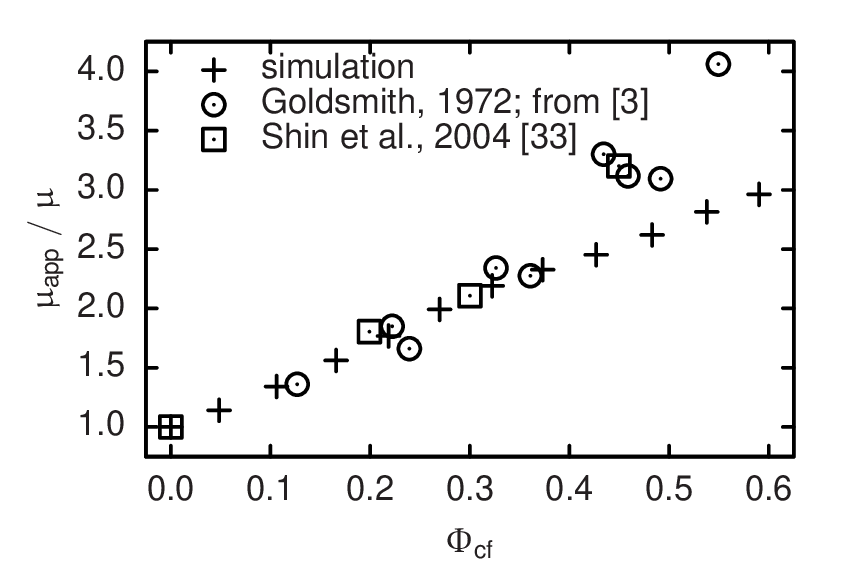}
  \caption{\label{fig:shear-volume-concentration}\RB{$\mu_\text{app}$
      dependence on} the volume concentration $\Phi_\text{cf}$ related
    to the cell-fluid interaction at a fixed shear rate of
    $\dot{\gamma}'=(2.2\pm0.1)\times10^3\,\text{s}^{-1}$ compared to
    experimental data for $\dot{\gamma}'>100\,\text{s}^{-1}$ given
    in~\cite{fung81,shin04}.}
\end{figure}

While the previous simulations regard bulk properties we now \RA{turn
  to} example\RA{s} where confinement and particulate effects play a
crucial role. \RA{The cell-wall interaction stiffness
  $\bar{\varepsilon}_\text{w}$ can be determined similarly as
  $\bar{\varepsilon}$ by comparison with experimental data. As an
  example we choose the sieving experiments performed by Chien et al.\
  who filtered human erythrocytes through polycarbonate sieves with
  mean pore diameters of $D'=2.2\,\mu\text{m}$ to $4.4\,\mu\text{m}$
  and mean pore lengths of $13\,\mu\text{m}$ \cite{chien71}. They
  analyzed the resulting flow resistance and damaging of cells in
  dependence on the pressure drop $\Delta P'$ across the sieves which
  was varied between approximately $10^2$ and
  $10^5\,\text{N}/\text{m}^2$. We simulate a single cell in front of a
  pore at a small value of $\Delta
  P'=4\times10^2\,\text{N}/\text{m}^2$ ($0.3\,\text{cm}\,\text{Hg}$)
  and vary the pore diameter and $\bar{\varepsilon}_\text{w}$. At this
  pressure drop, no significant hemolysis, which---as a sub-cell
  effect---is not resolved in our model, was found in the
  experiments~\cite{chien71}. However, compared to the case of
  $D'=4.4\,\mu\text{m}$, the flow resistance was increased by a factor
  of approximately $4$ for $D'=3.7\,\mu\text{m}$, by about $30$ for
  $D'=3.0\,\mu\text{m}$, and by more than $100$ for
  $D'=2.2\,\mu\text{m}$ at a hematocrit not higher than of the order
  of $1\,\%$. In our simulations, we identify the non-passing of model
  cells with a high increase of flow resistance in the experiments. We
  find that for
  $\bar{\varepsilon}'_\text{w}=1.47\times10^{-16}\,\text{J}$, the cell
  passes a pore with only $D'=3.0\,\mu\text{m}$ while for
  $\bar{\varepsilon}'_\text{w}=1.47\times10^{-14}\,\text{J}$ already a
  diameter of $D'=4.4\,\mu\text{m}$ proves an insurmountable
  obstacle. In view of reference~\cite{chien71} these two
  $\bar{\varepsilon}_\text{w}$ are unrealistic but the intermediate
  value
  $\bar{\varepsilon}'_\text{w}=\bar{\varepsilon}'=1.47\times10^{-15}\,\text{J}$
  is an appropriate choice for this setup which allows the model cell
  to pass through pores with a diameter of $3.7\,\mu\text{m}$ and
  more.} We \RA{now} study the flow through a bifurcation of
cylindrical capillaries with a radius of $4.7\,\mu\text{m}$. One of
the branches, however, features a stenosis with radius $R_\text{s}$. A
cut through the geometry containing nine RBCs is displayed in
\picref{junction-tiny-4}. It visualizes the cells as the approximated
ellipsoidal volumes defined by the zero-energy surface of the
cell-cell interaction and the vessel walls as midplane between fluid
and wall nodes. The open ends of the system are linked
periodically. We drive the system by means of a body force acting only
on the fluid in the entrance region. As initial condition, cells are
placed randomly in the unconstricted parts of the system. Both, the
tube diameters and Reynolds numbers $Re\lesssim4\times10^{-3}$ match
physiological situations \cite{fung81}. \RA{As above, the} cell-wall
potential stiffness is chosen \RA{to be}
$\bar{\varepsilon}'_\text{w}=\bar{\varepsilon}'=1.47\times10^{-15}\,\text{J}$. We
average the relative flow rate through the constricted branch
$\hat{Q}_\text{constr}=Q_\text{constr}/(Q_\text{constr}+Q_\text{unconstr})$
from $1.7\,\text{s}$ to $3.0\,\text{s}$ measured from system
initialization. This is done for $R'_\text{s}=2.7\,\mu\text{m}$,
$4.0\,\mu\text{m}$, and $4.7\,\mu\text{m}$. As expected, the results
in \picref{flow-tiny-ew0.05} are monotonous with $R_\text{s}$. When
studying the volumetric flow rates of plasma and cells separately, it
becomes clear that for $R'_\text{s}=2.7\,\mu\text{m}$ the cells cannot
pass the constriction \RA{at the present body force}. This situation
is visualized in \picref{junction-tiny-4}.

\begin{figure}
  \centering
  \includegraphics[width=\columnwidth]{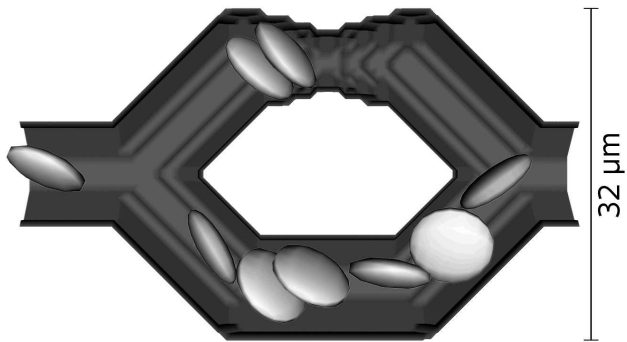}
  \caption{\label{fig:junction-tiny-4}Cut through a capillary
    bifurcation. Shown are the volumes defined by the cell-cell
    interaction and the midplane between wall and fluid nodes. The
    plasma is not visualized. The flow direction is from left to
    right. The vessel radius is $R'_\text{s}=2.7\,\mu\text{m}$ at a
    stenosis in the upper branch and $4.7\,\mu\text{m}$
    otherwise. Geometries with length scales that are not large
    compared to a cell diameter require treatment by a method that is
    able to resolve particulate effects like the shown clogging of the
    constricted branch for a cell-wall potential strength of
    $\bar{\varepsilon}'_\text{w}=1.47\times10^{-15}\,\text{J}$.}
\end{figure}

\begin{figure}
  \centering
  \includegraphics[width=\columnwidth]
  {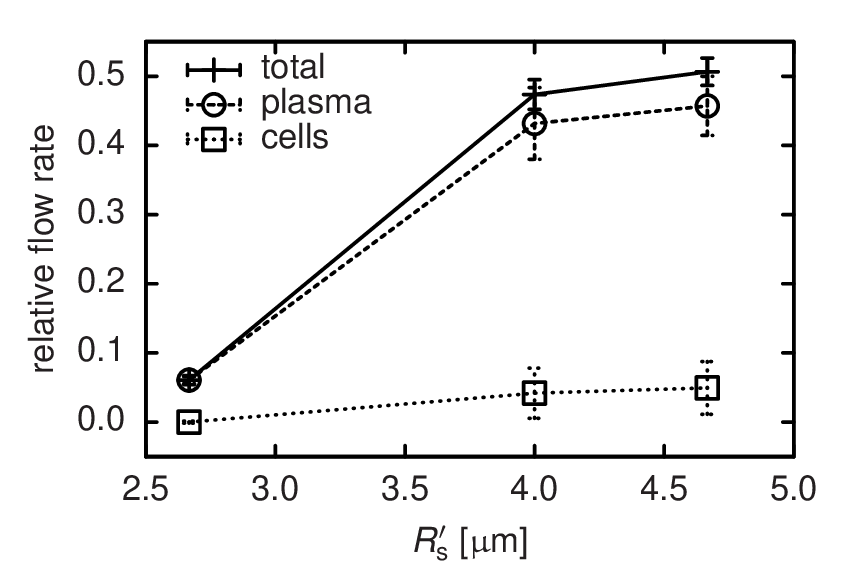}
  \caption{\label{fig:flow-tiny-ew0.05}Time-averaged relative flow
    rate through the constricted capillary in a bifurcation as shown
    in \picref{junction-tiny-4} for different stenosis radii
    $R_\text{s}$. The cell-wall interaction stiffness is
    $\bar{\varepsilon}'_\text{w}=1.47\times10^{-15}\,\text{J}$. While
    for $R'_\text{s}=2.7\,\mu\text{m}$, the constricted branch becomes
    clogged and only a small amount of plasma passes the remaining
    aperture, no clogging occurs for larger $R_\text{s}$.}
\end{figure}

\xRA{Even though a definitive conclusion is not possible without
  analyzing the pressure drop across the stenosis, this behavior seems
  to be unrealistic since in experiments RBCs were able to squeeze
  through significantly smaller pores~\cite{goldsmith75}.} We
\RA{further} study the dynamics of the system for the present and two
lower cell-wall interaction parameters and plot
$\hat{Q}_\text{constr}(t)$ in \picref{flow-tiny-r4}. For
$\bar{\varepsilon}'_\text{w}=1.47\times10^{-15}\,\text{J}$, the curve
decreases in two steps due to the successive arrival of two
erythrocytes and stays below $10\,\%$ as only a small amount of plasma
is able to pass the remaining aperture. In contrast, there is a
continuous flow of plasma and cells for
$\bar{\varepsilon}'_\text{w}=1.47\times10^{-17}\,\text{J}$. For an
intermediate stiffness
$\bar{\varepsilon}'_\text{w}=1.47\times10^{-16}\,\text{J}$ the cells
get stuck initially. However, the flow in the narrowed branch is
influenced by the time-dependent cell configuration in the other
branch. It happens eventually that the pressure in front of the
stenosis rises to a level which lets the RBC overcome the barrier
imposed by the cell-wall potential. Each restitution of a higher flow
level is initiated by a peak which can be explained by the cell-wall
potential that accelerates the RBC into the flow direction while the
cell leaves the constriction. As another effect, we find
$\bar{\varepsilon}_\text{w}$ to affect the relative flow rates of the
two phases since larger values force the cells into the center of the
capillaries where higher velocities are measured.

\begin{figure}
  \centering
  \includegraphics[width=\columnwidth]{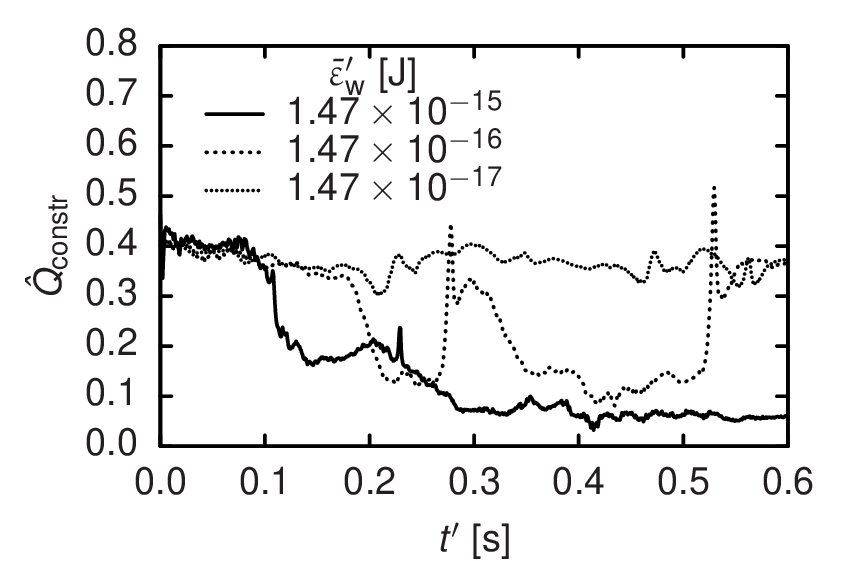}
  \caption{\label{fig:flow-tiny-r4}Time evolution of the relative flow
    rate through the constricted capillary in the bifurcation shown in
    \picref{junction-tiny-4}. For a cell-wall interaction stiffness of
    $\bar{\varepsilon}'_\text{w}=1.47\times10^{-15}\,\text{J}$, the
    relative flow rate drops to less than $10\,\%$ in two major steps
    related to the successive arrival of two single erythrocytes at
    the constriction. With
    $\bar{\varepsilon}'_\text{w}=1.47\times10^{-16}\,\text{J}$, the
    reduction of the flow rate is only temporary, since the cells are
    eventually able to pass. While leaving the constriction, the RBCs
    are accelerated by the cell-wall potential forces which explains
    the peaks of the flow rate.}
\end{figure}

Despite the coarse-graining of the model it qualitatively reproduces
some aspects of the behavior observed for blood flow in
capillaries. The fact that cells approaching a bifurcation show a
strong preference to choose the faster branch is described in
\cite{fung81}. The last consequence of this effect is visible in
\picref{junction-tiny-4} and \picref{flow-tiny-r4} where during a
considerable amount of time no further RBCs enter the constricted
branch after its closure. It can be seen that our model is able to
describe particulate effects which could hardly be covered in terms of
a continuous fluid. Obviously, reproducing the behavior of single
cells at bifurcations is crucial if the microcirculation and its
heterogeneous flow properties are to be modeled~\cite{popel05}.

The vessel radii present in the human microvascular system
approximately cover a range from $2\,\mu\text{m}$ to
$50\,\mu\text{m}$. Having demonstrated the applicability of our model
to small capillaries, we proceed with a study of the steady flow
through a larger vessel with a radius of $R'=31\,\mu\text{m}$
corresponding to an arteriole or venule~\cite{popel05}. In the
simulation, the vessel is closed periodically at a length of
$43\,\mu\text{m}$. We choose $\Phi_\text{cf}=42\,\%$ and an
intermediate cell-wall interaction stiffness
$\bar{\varepsilon}'_\text{w}=1.47\times10^{-16}\,\text{J}$. \picref{channel-h42-g1e-6}
shows a cut through the vessel \RB{for steady flow at $Re\sim1$}. The
flow is driven by a body force which acts on both plasma and cells in
the whole system and is equivalent to a constant macroscopic pressure
gradient. The system is evolved in time until neither the initial fcc
ordering of the cells nor significant directed changes in the
volumetric flow rate $Q$ are visible. In \picref{channel-u-profile},
the radial velocity profile in the case of a body force resulting in a
pseudo-shear rate of $\bar{v}'_z=Q/(2\pi
R^3)=1.3\times10^3\,\text{s}^{-1}$ \RB{or a Reynolds number of
  $Re\sim10$} is shown. The graph deviates from the parabolic
Hagen-Poiseuille profile that could be observed for a Newtonian
fluid. Instead a parabolic core region and a narrow boundary region
with high shear rates can be identified. The fit in
\picref{channel-u-profile} shows that this profile can be easily
explained by the modified axial-train model as described by
Secomb~\cite{secomb03}. The fit parameters are the viscosity ratio of
core and boundary $\mu_\text{c}/\mu_\text{b}$ and the width of the
cell-depleted boundary layer $\delta$. The obtained viscosity ratio of
$\mu_\text{c}/\mu_\text{b}=2.43\pm0.01$ is consistent with the bulk
properties in \picref{shear-volume-concentration} if we assume
$\mu_\text{b}=\mu$ and $0.4<\Phi_\text{cf}<0.5$ in the core. Also our
result of $\delta'=(1.47\pm0.04)\,\mu\text{m}$ seems compatible with
the value of $1.8\,\mu\text{m}$ suggested by
Secomb~\cite{secomb03}. In additional studies of radial cell-fluid
concentration profiles we prove the existence of a cell-depleted layer
and the possibility to tune its width by the cell-wall potential
stiffness $\bar{\varepsilon}_\text{w}$. \RA{We also find an increased
  cell concentration of up to around $\Phi_\text{cf}=60\,\%$ close to
  the central axis of the vessel. This must be a collective effect
  since in consistency with a 2D study by Qi et al.~\cite{qi02}, we
  observe that single cells in Poiseuille flow migrate to an
  intermediate lateral position between vessel wall and center.}

\begin{figure}
  \centering
  \includegraphics[width=\columnwidth]{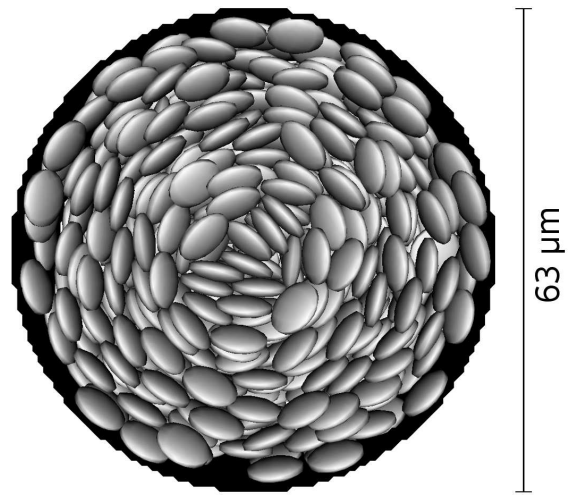}
  \caption{\label{fig:channel-h42-g1e-6}Cut through a cylindrical
    vessel with a radius of $31\,\mu\text{m}$. For this geometry we
    choose a cell-wall interaction strength of
    $\bar{\varepsilon}'_\text{w}=1.47\times10^{-16}\,\text{J}$. Shown
    are the volumes defined by the cell-cell interaction at $42\,\%$
    cell-fluid volume concentration and the midplane between wall and
    fluid nodes. The flow is pointing into the drawing plane and has a
    maximum velocity of $1.08\times10^{-2}\,\text{m}/\text{s}$ at the
    center.}
\end{figure}

\begin{figure}
  \centering
  \includegraphics[width=\columnwidth]
  {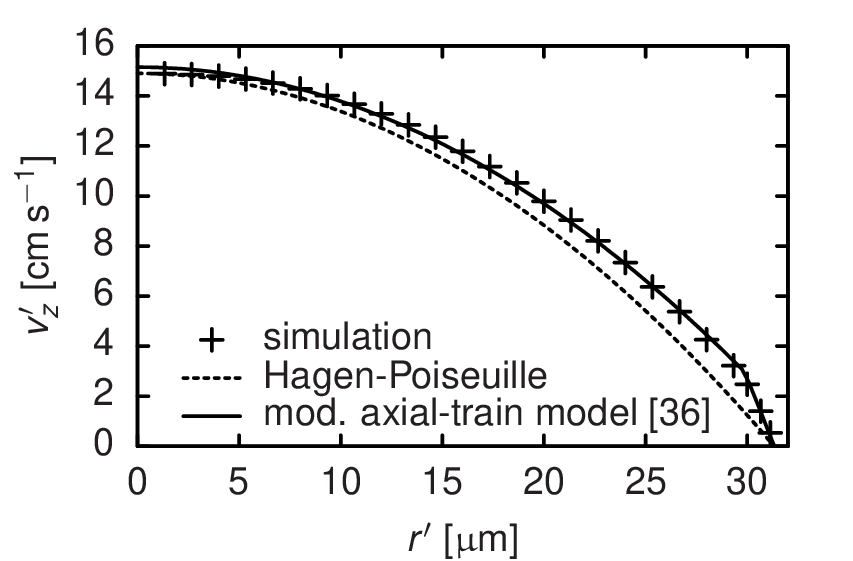}
  \caption{\label{fig:channel-u-profile}Radial velocity profile in a
    cylindrical vessel with $42\,\%$ average cell-fluid volume
    concentration. Apparent slip due to a cell depletion layer is
    visible. The profile can be well fitted by a modified axial-train
    model as described by Secomb~\cite{secomb03}. The parabolic
    Hagen-Poiseuille profile is plotted as well for comparison.}
\end{figure}

The apparent viscosity for a cylindrical vessel is calculated as
\begin{equation}
  \mu_\text{app}
  =
  \frac{\pi R^4}{8Q}
  \frac{\mathrm{d}P}{\mathrm{d}z}
\end{equation}
with $\mathrm{d}P/\mathrm{d}z$ being the macroscopic pressure
gradient~\cite{popel05}. Pries et al.~\cite{pries92b} present an
empirical expression for the dependency of $\mu_\text{app}$ on the
radius and hematocrit for the case of high flow rates after combining
a variety of experimental studies for pseudo-shear rates
$\bar{v}'_z>50\,\text{s}^{-1}$ in a single fit. We perform a series of
simulations at $R'=31\,\mu\text{m}$ and \RX{three} fixed pseudo-shear
\RX{rates between $\bar{v}'_z=Q/(2\pi R^3)=(62\pm1)\,\text{s}^{-1}$
  and $(563\pm3)\,\text{s}^{-1}$} but varying cell-fluid volume
concentrations $\Phi_\text{cf}$. \RB{The corresponding Reynolds
  numbers are $Re\lesssim1$.} If we identify $\Phi_\text{cf}$ with the
hematocrit as we do in \picref{shear-volume-concentration}, we
\RX{find very good agreement with} the relationship by Pries et
al.~\cite{pries92b}. The comparison is plotted in
\picref{channel-n-viscosity}.
\begin{figure}
  \centering
  \includegraphics[width=\columnwidth]
  {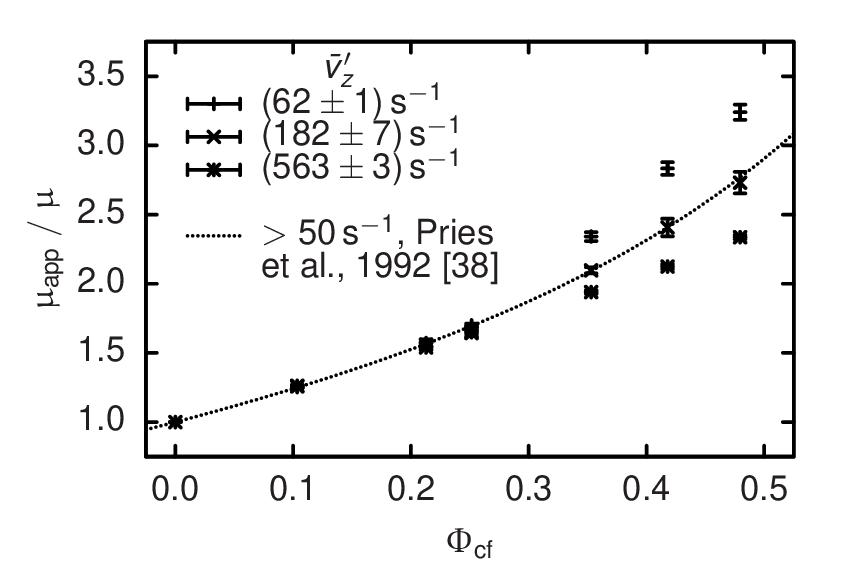}
  \caption{\label{fig:channel-n-viscosity}\RB{Dependence of the
      apparent dynamic viscosity $\mu_\text{app}$ in a cylindrical
      vessel with radius $R'=31\,\mu\text{m}$ on} the cell-fluid
    volume concentration $\Phi_\text{cf}$. \RX{Three} pseudo-shear
    \RX{rates $\bar{v}'_z=Q/(2\pi R^3)$ are examined}. The empirical
    result by Pries et al.~\cite{pries92b} for
    $\bar{v}'_z>50\,\text{s}^{-1}$ with $\Phi_\text{cf}$ as tube
    hematocrit is plotted for comparison.}
\end{figure}

The presence of a cell-depleted layer is closely connected to the
emergence of heterogeneous cell concentrations in different parts of
the microvasculature since branching daughter-vessels first of all
drain blood from the boundary layer~\cite{fung81}. The hematocrit, in
turn, influences the flow resistance, the flow rate, and the resulting
distribution of erythrocytes at branching points~\cite{popel05}. Our
simulation approach reproduces these aspects at least
qualitatively. When implemented together with an indexed LB scheme as
in \cite{axner08,mazzeo08,bernaschi09}, the method will be able to
simulate flow through digitized vessel networks covering the whole
scale of the microcirculation with high efficiency. Such simulations
are still computationally demanding despite the simplifications of the
model. Thus even systems that are small in physical units require
parallel supercomputers which makes the scalability of the code
crucial. For a quasi-homogeneous chunk of suspension consisting of
$1024^2\times2048$ lattice sites and $4.1\times10^6$ cells simulated
on a BlueGene/P system, we achieve a parallel efficiency normalized to
the case of $2048$-fold parallelism of $95.7\,\%$ on $16384$ and still
$85.2\,\%$ on $32768$ cores. In comparison, the pure LB code without
the MD routines that are responsible for the description of the cells
shows a relative parallel efficiency of $98.1\,\%$ on $32768$
cores. The parallel performance of the combined code is mainly limited
by the relation of the interaction range of a cell to the size of the
computational domain dedicated to each task. We are aware of only one
work on simulations of comparably large systems with a particulate
description of hemodynamics. This work was published by the group of
Aidun and models the deformation of cells explicitly
\cite{wu10,aidun10}. However, owing to the coarse-graining, our model
is easier to parallelize efficiently and---compared to the resolution
given in \cite{aidun10}---allows for substantially higher cell
numbers. Generally, our relatively low spatial resolution is highly
beneficial for the simulation of large systems since from
\eqnref{dt-constraint} it can be derived that the number of lattice
site updates necessary for the simulation of a system with a given
physical size for a given physical time interval scales with the fifth
power of $1/\delta x$. As for plain LB simulations, this number is a
good measure for the computational cost also in the case of our
suspension model.

\section{Conclusion}\label{ch:conclusion}

We developed a new approach for the coarse-grained simulation of
suspensions of soft particles. This approach is based on a
well-established method for rigid particle
suspensions~\cite{aidun98,ladd01} which covers the hydrodynamic
long-range interactions and phenomenological model potentials to
account for the behavior at small particle separations. A
parametrization suitable for the quantitative reproduction of
hemorheology at moderate to high shear
rates~\cite{chien70,fung81,shin04} was presented. \RA{The cell-wall
  interaction could be linked to experimental data on a single-cell
  level~\cite{chien71}.} Afterwards, we demonstrated that the model
shows a complex particulate behavior in bifurcations of partly
constricted capillaries which is an essential feature also of the flow
properties of the microcirculation in vivo~\cite{popel05}. Using the
example of steady flow through larger vessels, we proved the existence
of a cell-depleted layer and obtained radial velocity profiles that
are consistent with an accordant theoretical model~\cite{secomb03}. We
could even quantitatively reproduce the experimentally observed
dependency of the apparent viscosity in this geometry on the
hematocrit~\cite{pries92b}. These results suggest that following our
approach one can reproduce the particulate behavior of blood on a
range of spatial scales that up to this moment was not covered by a
single existing simulation method with comparable efficiency. Clearly,
our motivation is not to replace models with higher resolution like
the ones presented in \cite{sun06,dupin08,mcwhirter09,wu10}, but to
bridge the gap to continuous descriptions of blood. We believe that
this method can prove both an efficient tool for coarse grained yet
particulate simulations of flow in microvascular vessel networks and a
valuable contribution to the improvement of macroscopic blood
modeling.

\begin{acknowledgments}
  The authors thank A.\ C.\ B.\ Bogaerds and F.\ N.\ v.\,d.\ Vosse for
  fruitful discussions and S. Plimpton for providing his freely
  available MD code \textit{ljs} \cite{plimpton95}. Financial support
  is gratefully acknowledged from the Landesstiftung
  Baden-W\"urttemberg, the HPC-Europa2 project, the collaborative
  research centre (SFB) 716, and the TU/e High Potential Research
  Program. Further, the authors acknowledge computing resources from
  JSC J\"ulich, SSC Karlsruhe, \RX{CSC Espoo,} and SARA Amsterdam, the
  latter \RX{two} being granted by DEISA as part of the DECI-5
  project. \RX{We acknowledge the \textit{iCFDdatabase} for hosting
    the data (\textit{http://cfd.cineca.it}).}
\end{acknowledgments}

\end{document}